\documentclass[12pt,a4paper]{article}
\usepackage{graphicx}
\usepackage{times}
\newcommand\apj{{ApJ}}%
\newcommand\apjl{{ApJ}}%
\newcommand\aap{{A\&A}}%
\newcommand\mnras{{MNRAS}}%
\title{\bf The Orbit and Distance of WR\thinspace140}
\author{S. M.~Dougherty$^{1,2}$, V. Trenton$^{1,3}$, A. J.~Beasley$^4$\\
\vspace{1cm}\\
\normalsize $^1$ NRC-HIA DRAO, Penticton, BC, Canada\\
\normalsize $^2$ Institute for Space Imaging Science, University of Calgary, AB, Canada\\
\normalsize $^3$ University of Prince Edward Island, Charlottetown, PEI, Canada \\
\normalsize $^4$ National Ecological Observatory Network, Boulder, Colorado, USA 
\\
\\
\normalsize Published in proceedings of \\
\normalsize"Stellar Winds in Interaction", editors T. Eversberg and J.H. Knapen. \\ 
\normalsize Full proceedings volume is available on http://www.stsci.de/pdf/arrabida.pdf
}
\date{\mbox{}}
\begin{document}
\maketitle
%
%
\def\bull{\vrule height .9ex width .8ex depth -.1ex}
\makeatletter
\def\ps@plain{\let\@mkboth\gobbletwo
\def\@oddhead{}\def\@oddfoot{\hfil\tiny\bull\quad
Workshop ``Stellar Winds in Interaction'' Convento da Arr\'abida, 2010 May 29 - June 2\quad\bull}%
\def\@evenhead{}\let\@evenfoot\@oddfoot}
\makeatother
%
%
\def\beginrefer{\section*{References}%
\begin{quotation}\mbox{}\par}
\def\refer#1\par{{\setlength{\parindent}{-\leftmargin}\indent#1\par}}
\def\endrefer{\end{quotation}}
%
%
{\noindent\small{\bf Abstract:} A campaign of 35 epochs of
milli-arcsec resolution VLBA observations of the archetype
colliding-wind WR+O star binary system WR\thinspace140 show the
wind-collision region (WCR) as a bow-shaped arc of emission that
rotates as the highly eccentric orbit progresses.  The observations
comprise 21 epochs from the 1993-2001 orbit, discussed by Dougherty et
al. (2005), and 14 epochs from the 2001-2009 orbit, and span orbital
phase 0.43 to 0.95. Assuming the WCR is symmetric about the
line-of-centres of the two stars and ``points'' at the WR star, this
rotation shows the O star moving from SE to E of the WR star between
these orbital phases.  Using IR interferometry observations from IOTA
that resolve both stellar components at phase 0.297 in conjunction with
orbital parameters derived from radial velocity variations, the VLBA
observations constrain the inclination of the orbit plane as
$120^\circ\pm4^\circ$, the longitude of the ascending node as
$352^\circ\pm2^\circ$, and the orbit semi-major axis as
$9.0\pm0.1$~mas. This leads to a distance estimate to WR\thinspace140
of $1.81\pm0.08$~kpc. Further refinements of the orbit and distance
await more IR interferometric observations of the stellar components
directly. }
%
%
\section{Introduction}
The 7.9-year period WR+O system WR\thinspace140 (HD\thinspace193793)
is the archetype of colliding wind binary (CWB) systems. It is comprised of a WC7 star and an
O4-5 star in a highly elliptical orbit ($e\approx0.88$), where the
stellar separation varies between $\sim2$\,AU at periastron to
$\sim30$\,AU at apastron. This highly eccentric orbit clearly modulates
the dramatic variations observed in the emission from the system, from
X-ray energies to radio wavelengths (Williams et al.\ 1990). Perhaps
the most dramatic variations are observed at radio wavelengths, where
there is a slow rise from a low state close to periastron of a few
mJy, to a frequency-dependent peak in emission of 10's of mJy between
orbital phase 0.65 to 0.85, before a precipitous decline immediately
prior to periastron (see Fig.~1). A number of attempts to model these
variations have been made (e.g. Williams et al.\ 1990; White \&
Becker 1995) with limited success, though advances in our
understanding of WCRs are being made (e.g. Pittard \& Dougherty
2006; Pittard, these proceedings). Accurate orbital parameters are
critical inputs to these models.

\begin{figure}[h]
\centering \includegraphics[width=0.6\textwidth]{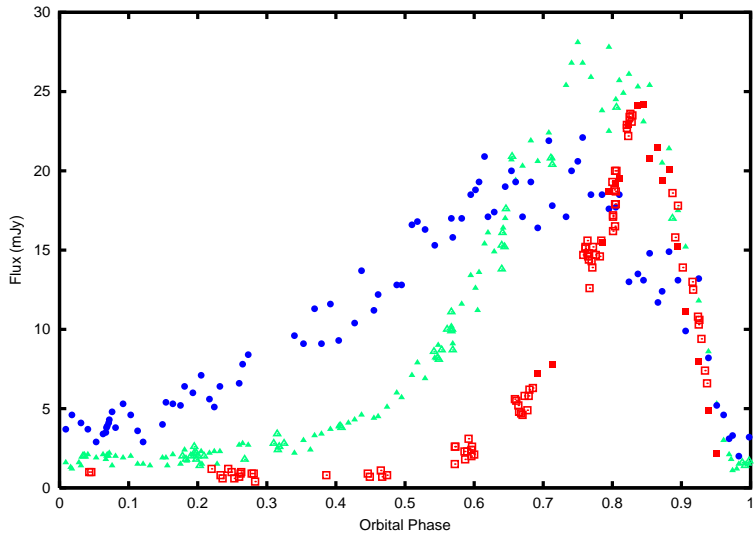}
\caption{Radio emission from WR\thinspace140 at 15 (blue circles), 5
  (green triangles), and 1.6\,GHz (red squares) as measured with the
  VLA (solid) and WSRT (open). Data taken from Williams et al. (1990)
  and White \& Becker (1995).\label{fig:flux_var}}
\end{figure}

Many of the orbital parameters in WR\thinspace140, in particular the
orbital period ($P$), epoch of periastron passage ($T_{\rm o}$),
eccentricity ($e$), and the argument of periastron ($\omega$) have
been established by others (see Marchenko et al.\ 2003 and references
therein), and refined most recently in an extensive observing campaign
during the 2009 periastron passage. However, the orbital inclination
($i$), semi-major axis ($a$), and the longitude of the ascending node
($\mathsf{\Omega}$) require the system to be resolved into a
``visual'' binary. The two stellar components in WR\thinspace140 have
been resolved using the Infra-red Optical Telescope Array (IOTA)
interferometer at a single epoch (Monnier et al.\ 2004). This single
observation sets the scale and orientation of the orbit since it
constrains potential families of solutions for $(i, a,
\mathsf{\Omega})$. Further epochs of IOTA observations have been
completed, but until analysis is complete, the VLBA observations of
the WCR offer the only means to determine the orbit direction and
constrain $i$, and hence $\mathsf{\Omega}$ and $a$.

An initial analysis of 21 epochs of VLBA observations taken between
1999 and 2000 (orbital phase 0.74 to 0.95) was described in
Dougherty et al. (2005). The work presented in this paper is an
amalgamation of those observations with an additional 14 observations
obtained between 2004 to 2008, that extended the orbital phase
coverage from 0.43 to 0.96. A re-analysis of the earlier observations
with the new data leads to tighter constraints on the derived orbit
parameters.

\section{VLBA observations of WR\thinspace140}

WR\thinspace140 was observed with the VLBA at 8.4 GHz at 35 epochs
between orbital phase 0.43 and 0.97. A selection of images are shown
in Fig.~2. The 8.4-GHz emission is clearly resolved, with a bow-shaped
emission region observed at many epochs. This shape is anticipated for
a WCR from model calculations (see, e.g., Eichler \& Usov 1993;
Dougherty et al.\ 2003; Pittard et al.\ 2006). The WCR rotates from
``pointing'' NNW to W over the observed orbital phases. This rotation
is key to determining the orbit of WR\thinspace140.  Assuming the arc
of emission is symmetric about the line-of-centres and points towards
the WR star, the O star is to the SSE of the WCR at epoch 0.43, and
approximately to the E at phase 0.96.

\begin{figure}[!ht]
\centering
\includegraphics[bb=569 27 42 764, angle=-90,width=0.45\textwidth]{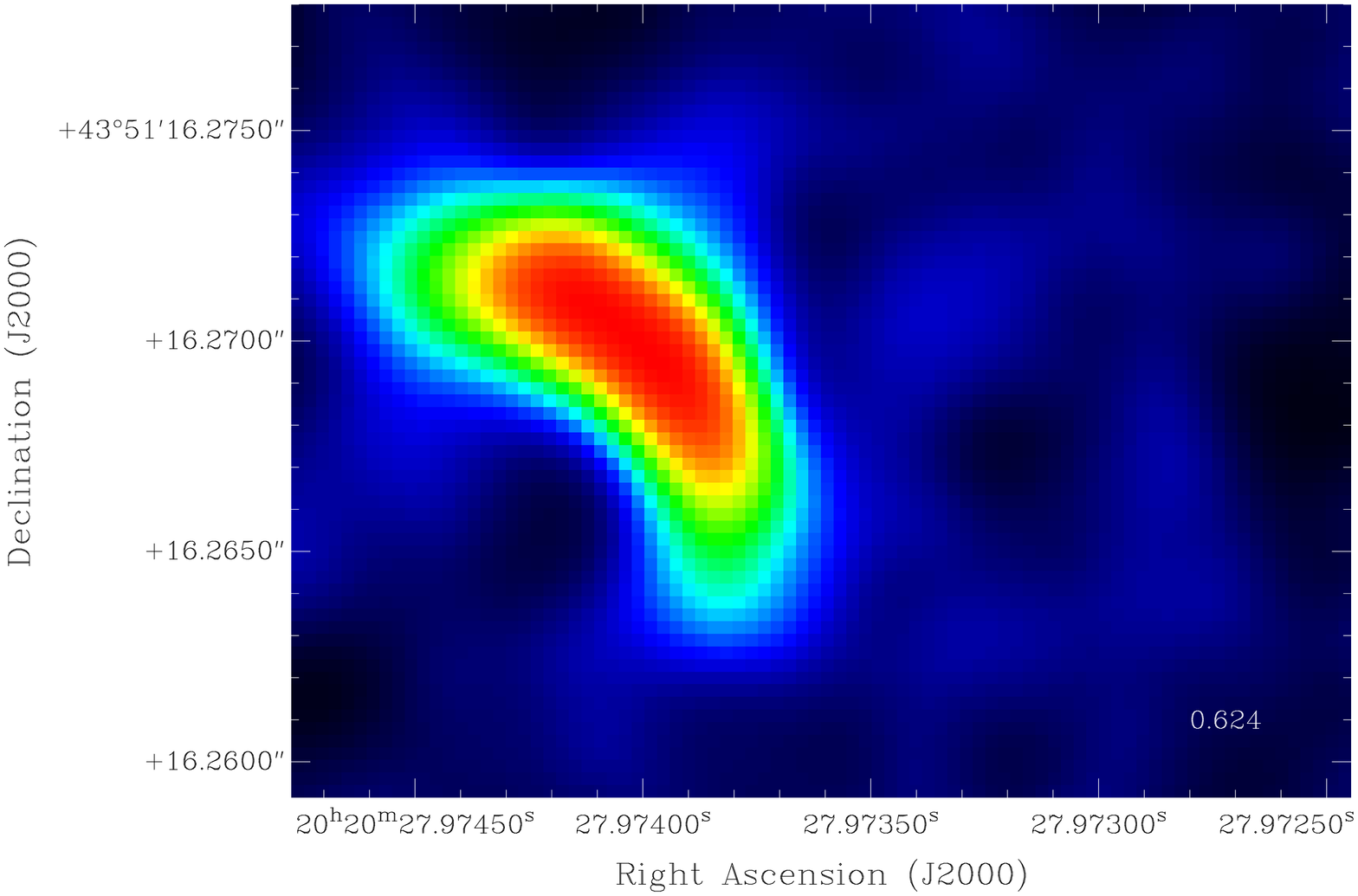}
\includegraphics[bb=29 76 538 449, height= 5.35cm, width=0.45\textwidth]{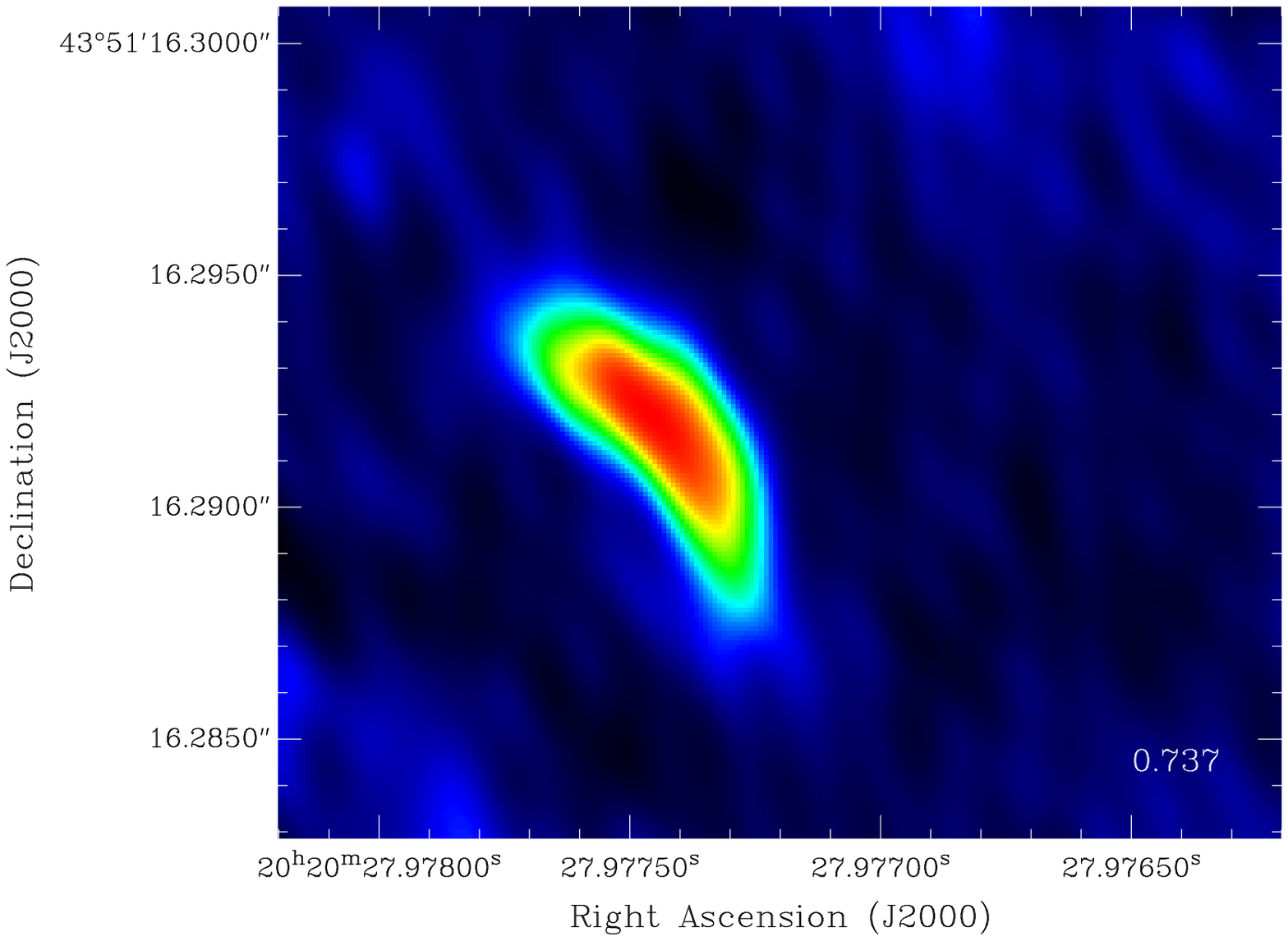}
\includegraphics[bb=569 27 42 764,angle=-90,width=0.45\textwidth]{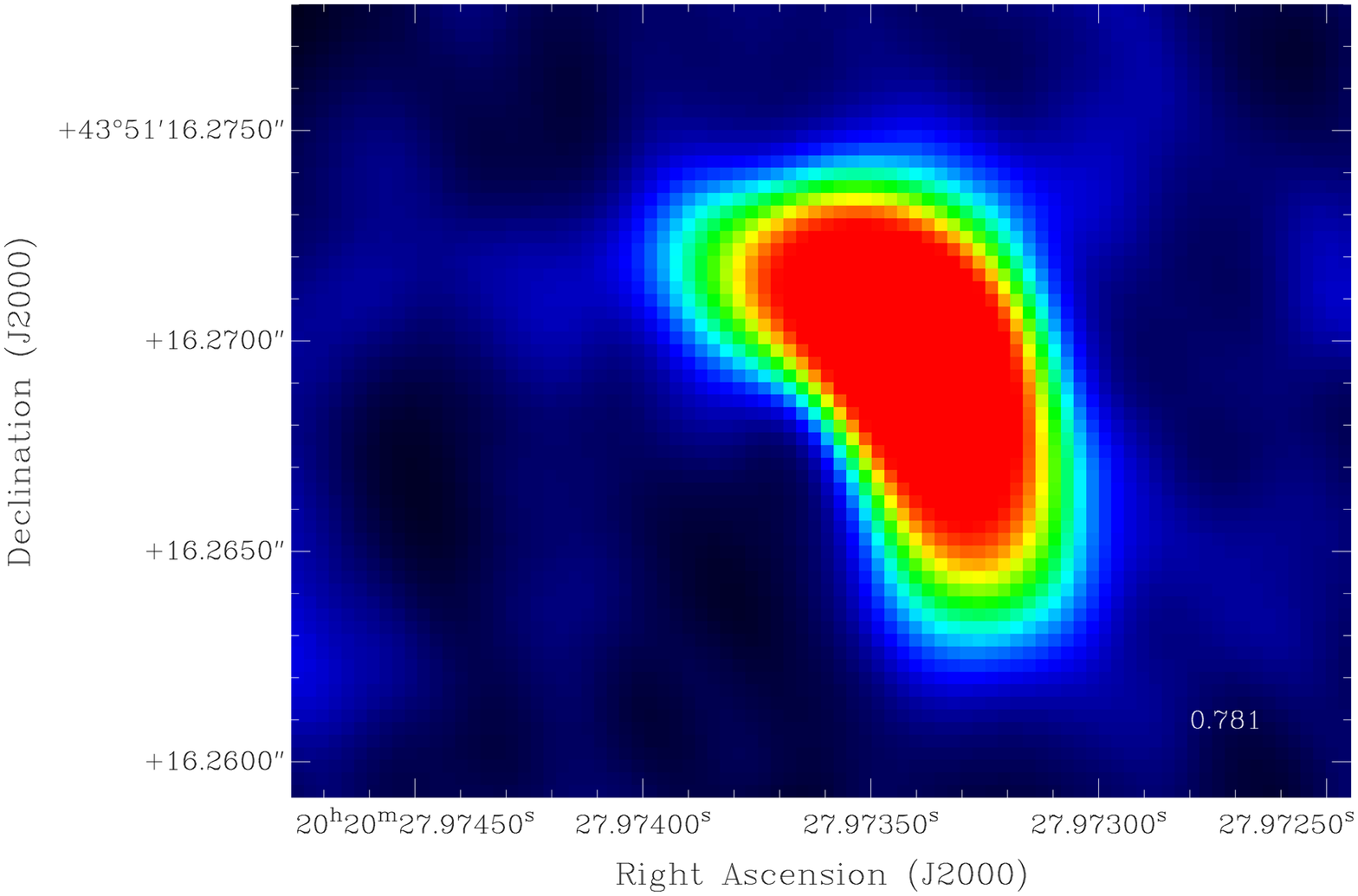}
\includegraphics[bb=29 76 538 449, height= 5.35cm, width=0.45\textwidth]{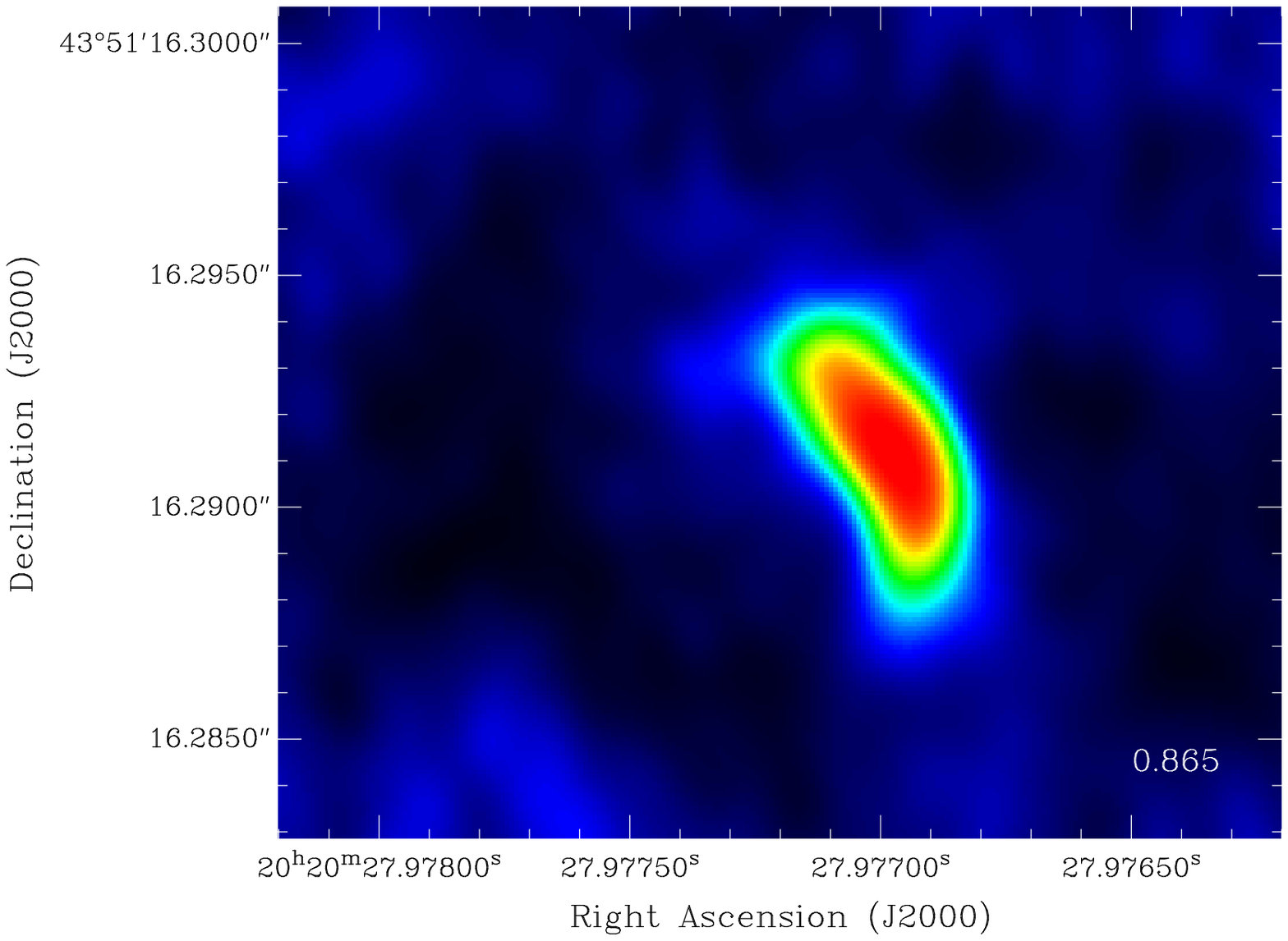}
\includegraphics[bb=569 27 42 764,angle=-90,width=0.45\textwidth]{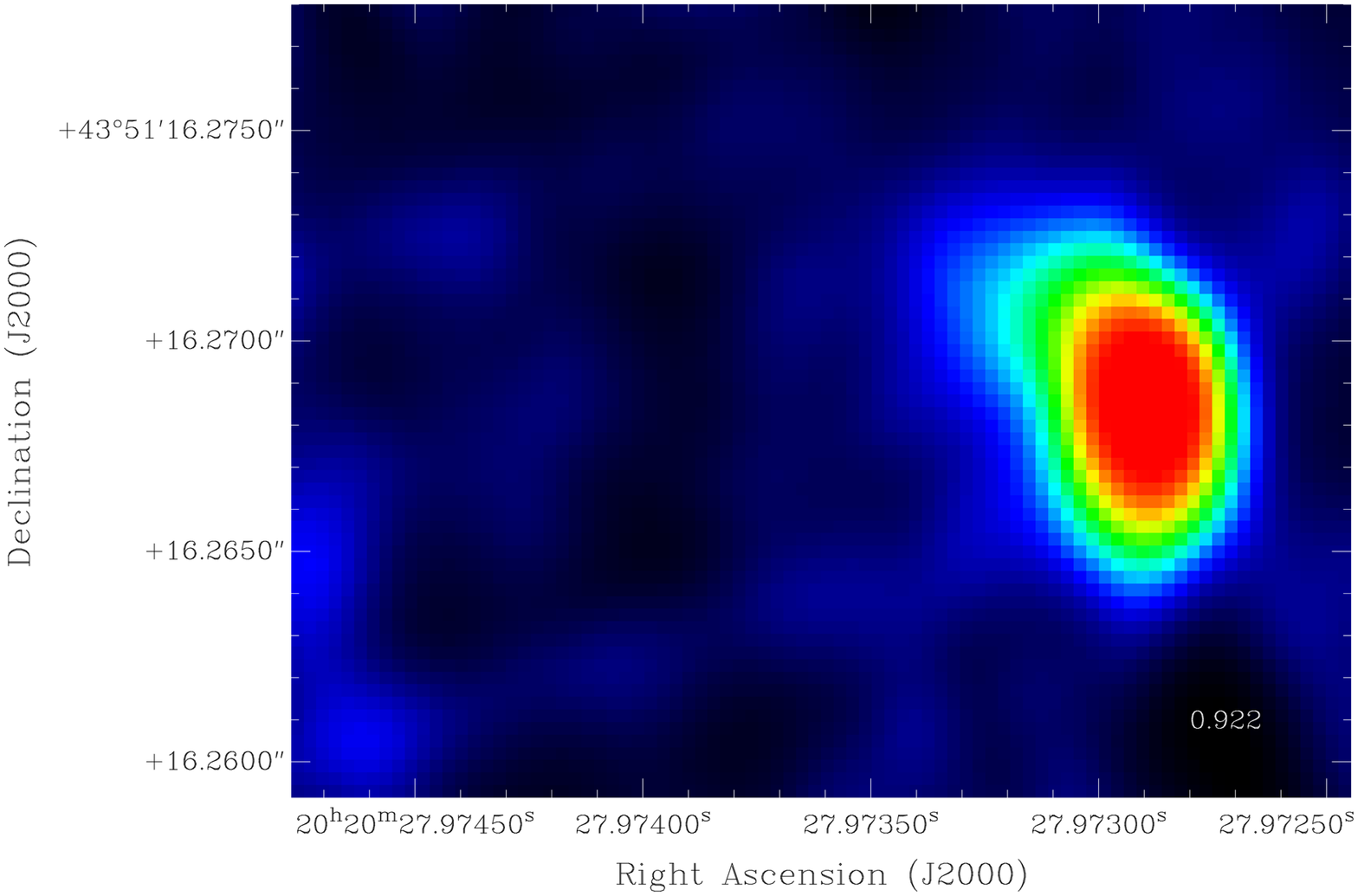}
\includegraphics[bb=29 76 538 449, height= 5.35cm, width=0.45\textwidth]{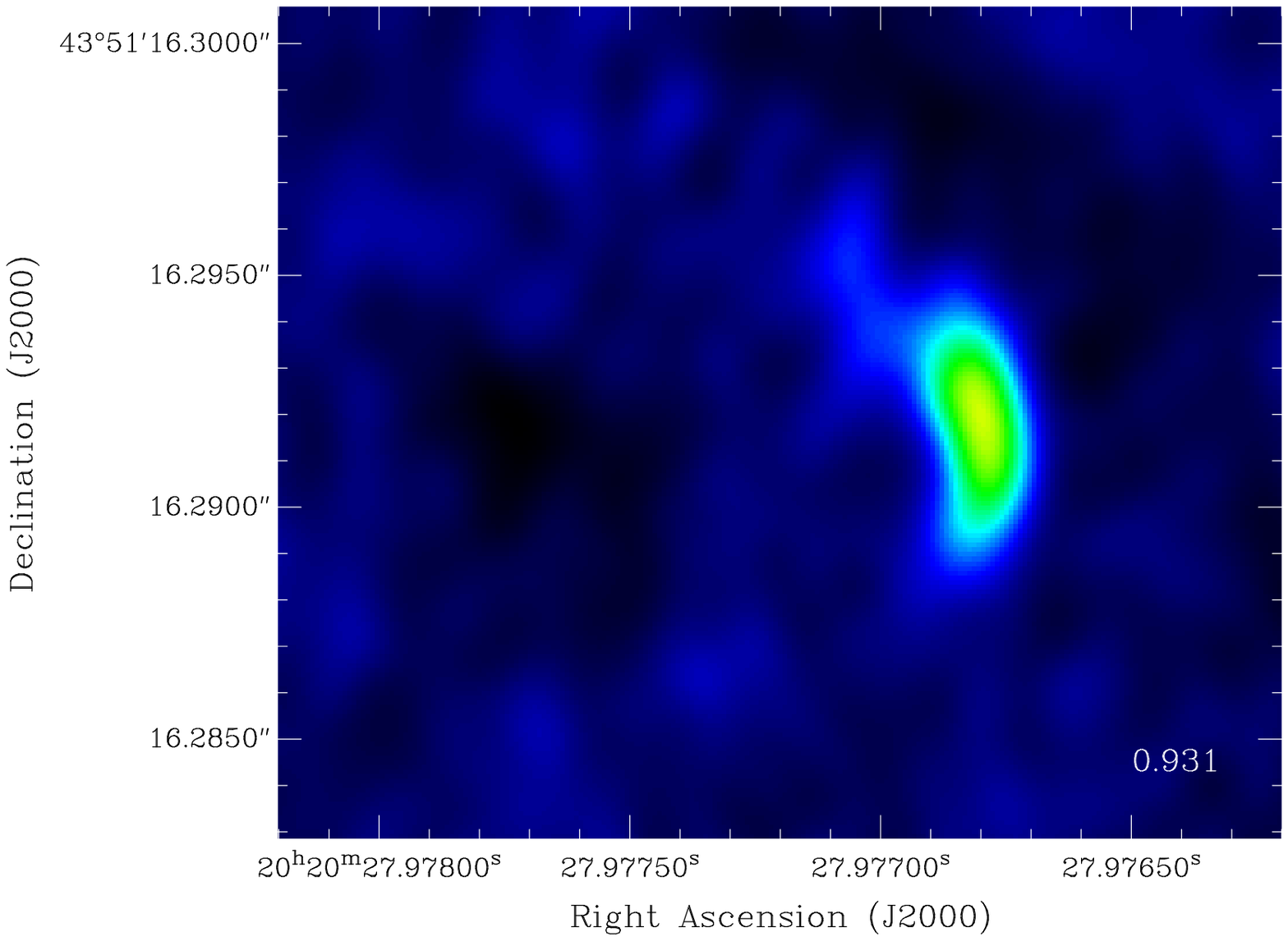}
\caption[]{VLBA 8.6-GHz images of WR140 at phases 0.62, 0.74, 0.78,
 0.87, 0.92, and 0.93 from the 1993-2001 orbit (phases 0.74, 0.87 and
 0.93---taken from Dougherty et al.\ 2005) and the 2001-2009
 orbit. The synthesized beam is 2.0$\times$1.5\,mas$^2$ in the
 1993-2001 orbit, and approximately $1.3\times$ that for the latest
 observations. Note change in RA and Dec between the images taken from
 the 1993-2001 orbit and the recent orbit. Rotation and proper motion
 of the WCR are evident during both orbits.  }
\end{figure}

\section{Deriving the Orbit}

On June 17, 2003 Monnier et al.\ (2004) observed WR\thinspace140 to
have a separation of $12.9^{+0.5}_{-0.4}$~mas at a position angle of
${151.7^{+1.8}_{-1.3}}$~degrees east of north.  Using $P=2896.6$~days,
$T_o=2446156.3$, $e=0.897$ and $\omega=46.8^\circ$, as determined from
analysis of observations during the 2009 periastron (Fahed et al.,
these proceedings), the observation at orbital phase 0.296 restricts
potential sets of solutions for $i, a$ and $\mathsf{\Omega}$ to those
shown in Fig.~3 for inclinations in the range $0^\circ<i<180^\circ$.

\begin{figure}[!ht]
\centering
   \includegraphics[width=0.6\textwidth,clip]{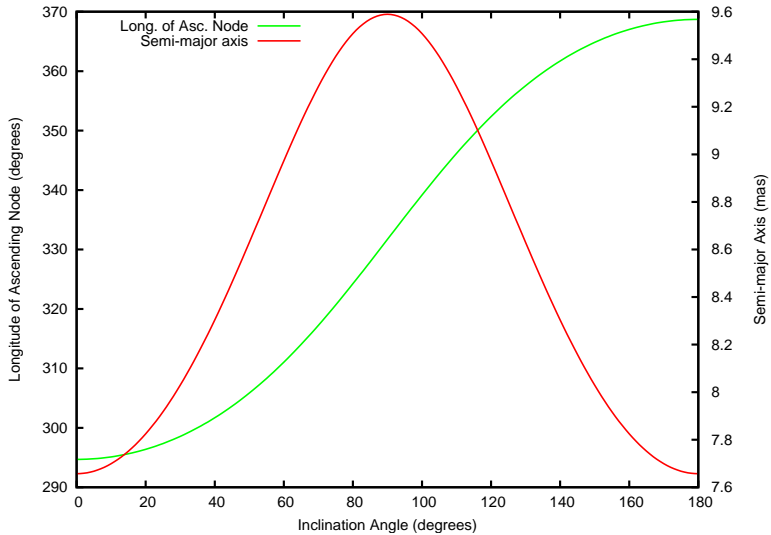}
\caption[]{Solutions for the longitude of the ascending node (red
line) and orbit semi-major axis (green line) as a function of orbit
inclination, derived from an IOTA separation and a position angle of
the stellar components at orbit phase 0.296 (Monnier et al.\ 2004). The
uncertainty in the IOTA observation gives an error in
$\mathsf{\Omega}$ of closely $\pm1^\circ$, and in the semi-major axis
of $\pm0.3$~mas.}
\end{figure}

The change in the orientation of the WCR with orbital phase gives the
inclination since each ($i,\mathsf{\Omega}$) solution family provides
a unique set of position angles as a function of orbital phase.  A
weighted minimum $\chi^2$ measure between the observed position angle
of the line of symmetry of the WCR, and by proxy the line-of-centres
of the stars, as a function of orbit phase determined for different
sets of ($i,\mathsf{\Omega}$) leads to a best-fit solution of
$i=120^\circ\pm4^\circ$ and $\mathsf{\Omega}=352^\circ\pm2^\circ$
(Fig.~4).  These values lead to a semi-major axis of
$a=8.97\pm0.13$~mas.

\begin{figure}[!ht]
\centering
   \includegraphics[width=0.6\textwidth,clip]{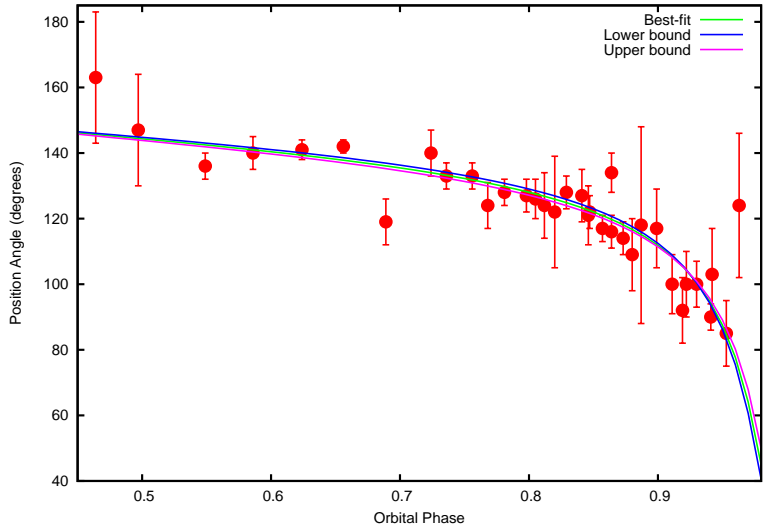}
\caption[]{The change in the position angle of the axis of symmetry of
the WCR as a function of orbital phase.  The green line is the
weighted best-fit curve of the position angle of the line-of-centres
of the two stars as a function of phase, corresponding to
$i=120^\circ$ and $\mathsf{\Omega}=352^\circ$. The other two lines
show the fits for $\mathsf{\Delta\chi}^2=\pm1$, showing the range of
potential fits for the quoted uncertainties.}
\end{figure}

\section{Distance of WR\thinspace140}

Stellar distance is one of the more difficult parameters to determine
for stars. Marchenko et al. (2003) determined $a\,\sin\,i =
14.10\pm0.54$~AU from radial velocity observations. From our estimate
of $i$, this gives $a=16.28\pm0.81$~AU. A semi-major axis of
$a=8.97\pm0.13$~mas then gives a distance estimate of
$1.81\pm0.09$~kpc, consistent with the previous estimate of
$1.85\pm0.16$~kpc by Dougherty et al. (2005). 

\section{Summary}

High-resolution radio observations of the WCR in WR140 have provided,
to date, the only way to determine the direction and orientation of
the orbit, starting from the scale of the orbit as deduced from IR
interferometry and orbital parameters from optical spectroscopy. From
this work it is possible to determine a precise, and hopefully
accurate, distance to WR140.

WR140 is the best colliding-wind binary system for attempts to
understand the underlying particle acceleration processes and physics
in wind-collision regions, in large part due to the wealth of
observational constraints. However, the orbit and distance are
critical to modelling the WCR, and further refinements of these
parameters await further IR interferometric observations that will
resolve the stellar components directly.
%
%
\section*{Acknowledgments}
Our thanks to Peredur Williams for presenting this work in our
absence during the workshop.  We are also grateful to Peredur for many
useful discussions related to this work. The work has been supported by
the National Research Council of Canada, and the University of Prince
Edward Island Co-op programme.  The observations presented here were
obtained from the Very Long Baseline Array, operated by the National
Radio Astronomy Observatory (NRAO).

%
%
\footnotesize
\beginrefer

\refer {Dougherty}, S.~M., {Beasley}, A.~J., {Claussen}, M.J., 
{Zauderer}, B.A., {Bolingbroke}, N.~J. 2005, \apj,~623, 447

\refer {Dougherty}, S.~M., {Pittard}, J.~M., {Kasian}, L., {Coker}, R.~F., {Williams},
  P.~M., \& {Lloyd}, H.~M. 2003, \aap,~409, 217.

\refer {Eichler}, D., \& {Usov}, V. 1993, \apj,~402, 271.

\refer {Marchenko}, S.~V., et~al. 2003, \apj,~596, 1295.

\refer {Monnier}, J.~D., et~al. 2004, \apjl,~602, L57.

\refer {Pittard}, J.~M., {Dougherty}, S.~M., {Coker}, R.~F., {O'Connor}, E. \& {Bolingbroke}, N.~J.
2006, \aap,~446, 1001.

\refer {Pittard}, J.~M. \& {Dougherty}, S.~M. 2006, \mnras,~372, 801.

\refer {White}, R.~L., \& {Becker}, R.~H. 1995, \apj,~451, 352.

\refer {Williams}, P.~M., {van der Hucht}, K.~A., {Pollock}, A.~M.~T., {Florkowski},
  D.~R., {van der Woerd}, H., \& {Wamsteker}, W.~M. 1990, \mnras,~243, 662.

\endrefer        

\newpage
\qquad
   
\end{document}